\documentclass[12pt]{article}
\usepackage{graphicx}

\newcommand{\beq}{\begin{equation}}
\newcommand{\eeq}{\end{equation}}
\newcommand{\beqn}{\begin{eqnarray}}
\newcommand{\eeqn}{\end{eqnarray}}
\newcommand{\beqns}{\begin{eqnarray*}}
\newcommand{\eeqns}{\end{eqnarray*}}

\begin{document}

\begin{titlepage}
\begin{center}

\hfill USTC-ICTS-19-13\\
\hfill May 2019

\vspace{2.5cm}

{\large {\bf Impact on the decay rate of $B_s\to \mu^+\mu^-$ from the dispersive two-photon transition}}\\
\vspace*{1.0cm}
 {Dao-Neng Gao$^\dagger$ \vspace*{0.3cm} \\
{\it\small Interdisciplinary Center for Theoretical Study,
University of Science and Technology of China, Hefei, Anhui 230026
China}}

\vspace*{1cm}
\end{center}
\begin{abstract}
\noindent
We study the long-distance contribution to $B_s\to\mu^+\mu^-$ decay, which is generated by the two-photon intermediate state via $B_s\to\gamma^*\gamma^*\to\mu^+\mu^-$ transition.
It is found that the dispersive two-photon amplitude can interfere with the dominant short-distance amplitude, which gives rise to new theoretical uncertainty in the branching ratio of $B_s\to\mu^+\mu^-$.
Our analysis shows that, by taking into account present experimental constraints, this uncertainty could be up to the same order of magnitude as some theoretical uncertainties of ${\cal B}(B_s\to\mu^+\mu^-)$ given in the past literature. Future precise studies of the double radiative $B_s\to\gamma\gamma$ decay, both experimentally and theoretically, may help to reduce the uncertainty. This novel effect has never been examined in $B_s\to\mu^+\mu^-$ decay.
\end{abstract}

\vfill
\noindent
$^{\dagger}$ E-mail:~gaodn@ustc.edu.cn
\end{titlepage}

\section{Introduction}

Rare leptonic $B$-meson decays $B_q\to \ell^+\ell^-$ with $q=d,\;s$ and $\ell=e,\;\mu,\; \tau$, which are helicity suppressed in the standard model (SM), could offer powerful tools to probe new physics scenarios beyond the SM.
Up to now, only the dimuon decay $B_s\to \mu^+\mu^-$ has been observed, and the first experimental evidence of this transition was reported by the LHCb Collaboration in 2012 \cite{LHCb2012}. Further observations with better signal significance were performed in Refs. \cite{LHCb2013, CMS}.  The most recent time-integrated branching ratio measurement by the LHCb experiment in 2017 \cite{LHCb2017} gives
\beq\label{LHCb-17}{\cal B}(B_s\to \mu^+\mu^-)=(3.0\pm 0.6^{+0.3}_{-0.2})\times 10^{-9},\eeq
and the current world average by the Particle Data Group \cite{PDG2018} is
\beq\label{worldaverage}
{\cal B}(B_s\to \mu^+\mu^-)=(2.7^{+0.6}_{-0.5})\times 10^{-9}.
\eeq
These measurements are in agreement with present SM predictions given in Refs. \cite{BGHMSS, BBS2018}. With higher experimental statistics, reduction of the experimental uncertainty will be expected in the future. It is thus important to increase the theoretical accuracy of the decay rate of $B_s\to \mu^+\mu^-$, which would eventually provide a precision test in flavor physics.

Theoretically, it is thought that the SM contributions to the $B_q\to\ell^+\ell^-$ decay can be described by an effective theory after integrating the heavy particles including the top quark, the Higgs boson, and weak gauge bosons $W$ and $Z$. The effective weak lagrangian relevant for the considered process, involving a single operator, reads \cite{BBL}
\beq\label{weaklagrangian}
{\cal L}_{\rm eff}={\cal N} {\cal C}_{10}{\cal Q}_{10}+ ...,
\eeq
where ${\cal Q}_{10}=(\bar{q}_L\gamma^\mu b_L)(\bar{\ell}\gamma_\mu \gamma_5\ell)$ and ${\cal C}_{10}$ is the Wilson coefficient. ${\cal N}$ is the normalization constant, containing some parameters such as the Fermi constant $G_F$ and the Cabibbo-Kobayashi-Maskawa (CKM) matrix elements etc., which will be shown explicitly below. The ellipses denote the sub-leading weak interaction terms. It is seen that the decay is characterized by a purely leptonic final state, its non-perturbative strong interaction effects are therefore confined to the matrix element
\beq\label{fbsq}
\langle 0|\bar{q}\gamma_\mu \gamma_5 b|\bar{B}_q(p)\rangle =i f_{B_q}p_\mu.
\eeq
Here the hadronic parameter $f_{B_q}$ is the $B_q$ decay constant, which can be computed in the framework of lattice QCD \cite{Aoki2017} with errors at a few percent level.
Thus the rare $B_q\to\ell^+\ell^-$ decay could be theoretically quite clean, which is indeed well suit for precision flavor physics.

In the SM, ${\cal B}(B_q\to \ell^+\ell^-)$ is proportional to the square of the Wilson coefficient ${\cal C}_{10}$ which can be computed within perturbation theory. The leading order contribution to ${\cal C}_{10}$ has been calculated for the first time by the authors of Ref. \cite{IN1981}, and the next-to-leading order(NLO) QCD corrections have been given in Refs. \cite{BB93-1, BB93-2, MU99, BB99}. Theoretical accuracy can be further improved by including the higher order corrections \cite{BGGI}. Recently, the NLO electroweak (EW) corrections and QCD corrections up to the next-to-next-to-leading order (NNLO) have been computed in Ref. \cite{BGS} and Ref. \cite{HMS}, respectively.
Interestingly, these two new calculations of the NLO EW and NNLO QCD corrections to  ${\cal C}_{10}$ were combined in the analysis of $B_q\to \ell^+\ell^-$ \cite{BGHMSS},
and the SM prediction for the muonic decay has been given by
\beq\label{BGHMSS}
{\cal B}(B_s\to \mu^+\mu^-)_{\rm SM}=(3.65\pm 0.23)\times 10^{-9}.
\eeq

As discussed in Ref. \cite{BGHMSS}, the dominant uncertainties of the theoretical prediction (\ref{BGHMSS}) are due to some parameters appearing in the calculation of the branching ratio: $4\%$ from the decay constant $f_{B_s}$, $4.3\%$ from CKM matrix elements, and $1.6\%$ from the top quark mass; while the nonparametric uncertainties, which are due to the omission of higher order QCD and electroweak corrections, as well as higher dimensional operators in the weak effective lagrangian, have been  significantly reduced to be at the level of around $1.5\%$, thanks to two new results on the NLO EW \cite{BGS} and NNLO QCD \cite{HMS} computations. Further reduction of the larger parametric uncertainties of ${\cal B}(B_s\to \mu^+\mu^-)$ will depend on the future improvement of the lattice determination of $f_{B_s}$ and measurement of SM parameters.

Very recently, it has been pointed out by the authors of Ref. \cite{BBS2018} that there exists a power-enhanced NLO electromagnetic correction to the $B_q\to\ell^+\ell^-$ decay, which, neglected in Ref. \cite{BGHMSS}, is due to the virtual photon exchanged between the final-state leptons and the light spectator antiquark $\bar{q}$ in the $B_q$ meson. These authors have found that the power-enhancement is directly related to the interplay of hard-collinear and collinear scales in the frame work of soft-collinear effective theory \cite{BS98, BFPS2001, BCDF2002}, and the impact of this effect on the branching ratio of $B_s\to\mu^+\mu^-$ is about $1\%$, of the same order of the nonparametric theoretical uncertainty in eq. (\ref{BGHMSS}). After taking into account this new correction, the SM prediction can be updated to \cite{BBS2018}
\beq\label{BBS}
{\cal B}(B_s\to \mu^+\mu^-)_{\rm SM}=(3.57\pm 0.17)\times 10^{-9}.
\eeq

\begin{figure}[t]
\begin{center}
\includegraphics[width=5cm,height=3cm]{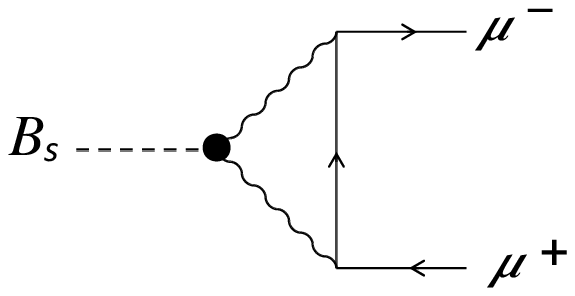}
\end{center}
\caption{The diagram that gives the transition $B_s\to\gamma^*\gamma^*\to \mu^+\mu^-$ with the wave line denoting the (virtual) photon, and the solid circle denotes some hadronic form factors.}\label{figure1}
\end{figure}

In this paper, we report on an investigation of another new correction to this muonic decay, which will be generated by the two-photon intermediate state via  the long-distance $B_s\to\gamma^*\gamma^*\to \mu^+\mu^-$ transition, as depicted in Fig. \ref{figure1}. The amplitude of this transition could be decomposed into the absorptive part given by the on-shell two-photon exchange, and the dispersive part contributed by the off-shell photons. The former part will be fixed once the amplitude of the double radiative $B_s\to \gamma\gamma$ decay is determined while the latter part, sensitive to the hadronic $B_s \gamma^*\gamma^*$ form factor, cannot be computed using the  model-independent approach. The similar study has been done in the neutral Kaon decay $K_L\to \mu^+\mu^-$, and it is found that the absorptive part by the two-photon cut provides the dominant contribution to its total decay rate \cite{LMS69,MRS70,DIP98,DP98,GV98,EKP,KPPR,IU04}. In our case, it will be not surprising that calculation of Fig. \ref{figure1} yields a small contribution to the branching ratio of $B_s\to \mu^+\mu^-$ since it is believed that, comparing eq. (\ref{LHCb-17}) with eqs. (\ref{BGHMSS}) and (\ref{BBS}), the short-distance amplitude given by eq. (\ref{weaklagrangian}) should play the dominant role in the leptonic $B$-meson decays. However, the small dispersive two-photon amplitude could interfere with the short-distance contribution, which might lead to some interesting effects on ${\cal B}(B_s\to \mu^+\mu^-)$.  It is of importance to estimate the possible theoretical uncertainty of the decay rate due to these corrections. This is the main purpose of the present paper.

\section{$B_s\to \gamma^*\gamma^*\to\mu^+\mu^-$ and its impact on ${\cal B}(B_s\to\mu^+\mu^-$)}

The general decay amplitude for $B_s\to\gamma\gamma$ can be parameterized as
\beq\label{Bsgammagamma}
{\cal A}(B_s\to \gamma\gamma)=\frac{G_F}{\sqrt{2}}f_{B_s}\langle \gamma\gamma| A_- F_{\mu\nu}\tilde{F}^{\mu\nu}+A_+ F_{\mu\nu}{F}^{\mu\nu}|0\rangle,
\eeq
where $F^{\mu\nu}$ is the photon field strength tensor, and $\tilde{F}^{\mu\nu}=1/2~\varepsilon^{\mu\nu\alpha\beta}F_{\alpha\beta}$ is its dual.The subscripts $\pm$ on $A_\pm$ denote the CP properties of the corresponding two-photon final states. We then obtain for the decay rate
\beq\label{rate-bs2gamma}
\Gamma(B_s\to\gamma\gamma)=\frac{G_F^2m_{B_s}^3f_{B_s}^2}{32\pi}\left(|A_-|^2+|A_+|^2\right).\eeq
Experimentally, this process has been not observed yet, and the present upper limit given by the Belle Collaboration \cite{Belle2015} is
\beq\label{Belle2015}
{\cal B}(B_s\to\gamma\gamma)< 3.1 \times 10^{-6}
\eeq
at the 90\% confidence level. We thus have $\sqrt{|A_-|^2+|A_+|^2}< 3.4\times 10^{-4}$. If the quantities $A_\pm$ are of the same order of magnitude, one has
\beq\label{A-}
|A_-|\sim |A_+|< 2.4 \times 10^{-4}.
\eeq
If $|A_-|\gg |A_+|$, we get
\beq\label{A-1}
|A_-|<3.4\times 10^{-4}.
\eeq

On the theoretical side, the double radiative $B_s$ decay has been studied extensively in the SM, in which the quark-level short-distance contributions with/without QCD corrections were calculated in Refs. \cite{LLY90,HK92,RRS97,CLY97,BB02, MG12}, and the long-distance contributions from the hadronic intermediate states were estimated in Refs. \cite {HI97, CE98, LZZ99}. The branching ratio of this mode was predicted, still with some large uncertainty, to be in the range of $10^{-7}\sim 10^{-6}$, below the current experimental upper limit in eq. (\ref{Belle2015}).

 Note that, from Fig. \ref{figure1}, the CP-even $A_+$ part amplitude in eq. (\ref{Bsgammagamma}) will lead to the scalar $\bar{\ell}\ell$ term while the CP-odd $A_-$ part will give rise to the $\bar{\ell}\gamma_5\ell$ structure for the leptonic decay. Therefore, we shall be not concerned about the $A_+$ part because it only generates a tiny contribution, which does not interfere with the dominant pseudoscalar short-distance amplitude given by eq. (\ref{weaklagrangian}). This is also the reason that we will not consider the $|A_-|\ll |A_+|$ case in the present study. Actually, theoretical calculations seems to support that they are of the same magnitude, for examples, as shown in Ref. \cite{BB02} for the short-distance contribution, and in Ref. \cite{LZZ99} for the long-distance contribution. Nevertheless, in our following numerical analysis, we still discuss the case of $|A_-|\gg |A_+|$ in order to show the possible largest uncertainties from the dispersive two-photon transition might be reached.

 Now it is straightforward to derive the amplitude of $B_s\to\mu^+\mu^-$ contributed by the two-photon intermediate state, focusing only on the $A_-$ part, which reads
 \beq\label{2gammmacontribution}
 i{\cal A}_{\gamma\gamma}=\frac{4 G_F f_{B_s} m_\mu}{\sqrt{2}}\frac{\alpha_{\rm em}}{4\pi}\bar{u}(q_-)\gamma_5 v(q_+)\cdot {I} \cdot A_-
 \eeq
 with
 \beq\label{R}
 I=\frac{2i}{\pi^2 m_{B_s}^2}\int {d^4k}\frac{k^2 p^2- (k\cdot p)^2}{k^2 (p-k)^2(\ell^2-m_\mu^2)} f(k^2, (p-k)^2).
 \eeq
Here $p^2=m_{B_s}^2$, $\ell=k-q_+$, and $q_+^2=q_-^2=m_\mu^2$. The function $f(k^2, (p-k)^2)$ is introduced to parameterize the hadronic $B_s\gamma^*\gamma^*$ form factor and normalized as $f(0,0)=1$. Considering this part contribution to the decay rate only, we have
\beq\label{ratioto2gamma}
\frac{{\cal B}(B_s\to\gamma^*\gamma^*\to \mu^+\mu^-)}{{\cal B}(B_s\to\gamma\gamma)}= \frac{2\alpha_{\rm em}^2r_\mu \beta_\mu}{\pi^2}\frac{|A_-|^2}{|A_-|^2+|A_+|^2}|{\cal I}|^2,
\eeq
where $r_\mu=m_\mu^2/m_{B_s}^2$ and $\beta_\mu=\sqrt{1-4 r_\mu}$.  As mentioned above, the absorptive part of the amplitude (\ref{2gammmacontribution}) for the on-shell two-photon intermediate state can be determined uniquely. In this case, the imaginary part of the integral ${I}$ is fixed as
\beq\label{ImI}
{\rm Im}~{I}=\frac{\pi}{2\beta_\mu}\log \frac{1-\beta_\mu}{1+\beta_\mu}=-12.35 \eeq
by using the experimental values of $m_\mu$ and $m_{B_s}$ \cite{PDG2018}. Consequently, one has
\beq\label{ratio-abs}
\frac{{\cal B}(B_s\to\gamma^*\gamma^*\to \mu^+\mu^-)_{\rm abs}}{{\cal B}(B_s\to\gamma\gamma)}=6.8\times 10^{-7}\frac{|A_-|^2}{|A_-|^2+|A_+|^2}.
\eeq
From the present upper limit shown in eq. (\ref{Belle2015}), we then obtain
\beq\label{abslimit}
{\cal B}(B_s\to\gamma^*\gamma^*\to \mu^+\mu^-)_{\rm abs}<2.1\times 10^{-12},
\eeq
which is very small and below $0.1\%$ of the dominant short-distance contribution given in eq. (\ref{BGHMSS}) or eq. (\ref{BBS}). This is very different from the $K_L$ case in which the absorptive part of $K_L\to\gamma\gamma\to\mu^+\mu^-$ almost saturates the experimental rate of $K_L\to \mu^+\mu^-$ \cite{IU04}. However, this does not mean that the effects induced from Fig. \ref{figure1} should be completely negligible since its dispersive part amplitude, although it may be also small, can interfere with the dominant short-distance amplitude, which would give rise to the significant impact on the decay rate of $B_s\to\mu^+\mu^-$.

By contrast with the absorptive part amplitude, to evaluate the dispersive two-photon contribution it is insufficient to know the on-shell $B_s\to\gamma\gamma$ amplitude. Unfortunately, the off-shell form factor $f(k^2, (p-k)^2)$, which is related to the long-distance hadronic physics, cannot be computed in a model-independent way. This situation will not change before we are able to calculate reliably the long-distance amplitude from QCD.
On the other hand, it is easy to see that, the integral ${I}$ in eq. (\ref{R}) will be logarithmically divergent when we turn off the form factor. Therefore, at present we have to employ the phenomenological parametrization
for the form factor to soften the ultraviolet divergence of the transition, in order to estimate the contribution of the dispersive two-photon amplitude. Due to Bose symmetry, the form factor function $f(k_1^2, k_2^2)$ should be symmetric under the interchange $k_1\leftrightarrow k_2$. As a simple realization to satisfy these requirements, one may take
\beq\label{formfactor1}
f(k_1^2,k_2^2)=\frac{1}{2}\left(\frac{M^2}{M^2-k_1^2}+\frac{M^2}{M^2-k_2^2}\right)
\eeq
or
\beq\label{formfactor2}
f(k_1^2,k_2^2)=\frac{M^4}{(M^2-k_1^2)(M^2-k_2^2)}.
\eeq
Here $M$ is thought of as the relevant cutoff, and we keep $M>m_{B_s}$ to avoid changing the absorptive part amplitude. Using these realizations, the long-distance two-photon contribution to $B_s\to\mu^+\mu^-$ is finite and can be computed in terms of $M$. The calculation is very standard. Explicitly, for the form factor (\ref{formfactor1}), we have
\beqn\label{ReI}
{\rm Re}~{I}=\frac{1}{\beta_\mu}\left[ {\rm Li}_2 \left(\frac{\beta_\mu-1}{\beta_\mu+1}\right)+\frac{\pi^2}{12}+\frac{1}{4}\log^2{\frac{1-\beta_\mu}{1+\beta_\mu}}\right]-\frac{7}{2}
                 - 3 g_1 (M) + \frac{1}{2} g_2(M),
\eeqn
where the dilogarithm function ${\rm Li}_2(x)=-\int^x_0 dt \log(1-t)/t$, and
\beqn\label{g1g2}
&&g_1(M)=\int^1_0 dx\int^{1-x}_0 dy~\log\left[r_\mu (1-x-y)^2- x y+r_M x\right],\\\nonumber\\
&&g_2(M)=\int^1_0 dx\int^{1-x}_0 dy~\frac{(1-4r_\mu)(1-x-y)^2}{r_\mu (1-x-y)^2- x y+r_M x}
\eeqn
with $r_M=M^2/m_{B_s}^2$.

\begin{figure}[t]
\begin{center}
\includegraphics[width=11cm,height=8cm]{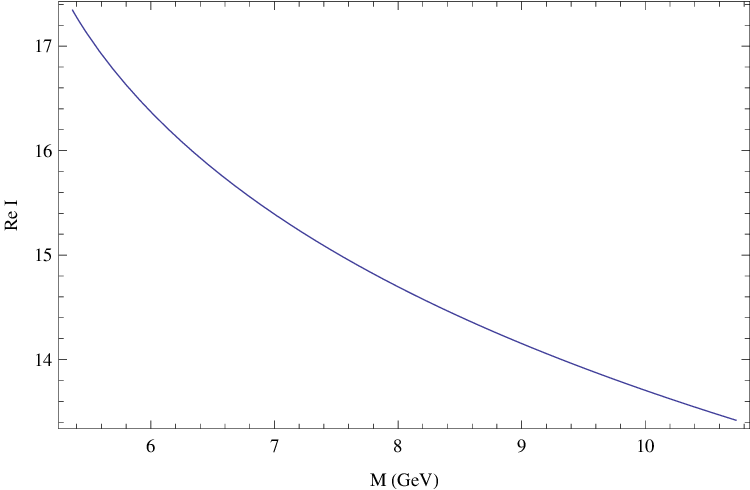}
\end{center}
\caption{${\rm Re}~I$ as a function of $M$ using the form factor of eq. (\ref{formfactor1}). }\label{figure2}
\end{figure}

Obviously, the functions $g_1(M)$ and $g_2 (M)$ can be integrated numerically for the fixed value of $M$. In order to evaluate the long-distance contribution to this muonic $B$ decay, it is reasonable to set $m_{B_s}<M<2 m_{B_s}$. Direct calculation thus shows that ${\rm Re}~{I}$ is in the range of $13.4 \sim 17.3$, not strongly dependent of the cutoff $M$, as displayed in Fig. \ref{figure2}. Similar analysis can be done using the form factor of eq. (\ref{formfactor2}), and ${\rm Re}~{I}$ will be from $15.3$ to $20.8$ for the same range of $M$. This is actually not very surprising since, after turning off the form factor, the integral $I$ in eq. (\ref{R}) contains only logarithmic divergence, which is in general not very sensitive to the cutoff. It is natural to expect that the dispersive part contribution is comparable in order of magnitude to the absorptive part. Comparing with eq. (\ref{ImI}), this is indeed the case in our calculation. Meanwhile, from eq.(\ref{ratioto2gamma}), it is seen that both the dispersive and absorptive parts will give tiny contributions to $B_s\to\mu^+\mu^-$ if we do not consider the interference with the dominant short-distance amplitude. In what follows we will estimate the interference effect by adopting
\beq\label{ReI1} {\rm Re}~{I} = 13.4 \sim 20.8. \eeq

The short-distance $\bar{B}_s\to \mu^+\mu^-$ decay amplitude can be expressed as \cite{BBS2018}
\beq\label{leadingorder} i{\cal A}=m_\mu f_{B_s}{\cal N}{\cal C}_{10}\bar{u}(q_-)\gamma_5 v(q_+) \eeq
with
\beq\label{N}
{\cal N}=V_{tb}V_{ts}^*\frac{4G_F}{\sqrt{2}}\frac{\alpha_{\rm em}}{4\pi}.
\eeq
This gives the decay rate for $B_s\to\mu^+\mu^-$ as
\beq\label{shortdistance-rate}
\frac{m^3_{B_s}f^2_{B_s}}{8\pi}|{\cal N}|^2 r_\mu \beta_\mu |{\cal C}_{10}|^2.
\eeq
To include the dispersive long-distance two-photon contribution of eq. (\ref{2gammmacontribution}), one can make the substitution
\beq\label{substitution}
{\cal C}_{10}\rightarrow {\cal C}_{10}+ \frac{A_-\cdot{\rm Re}~I}{V_{tb}^*V_{ts}}.
\eeq

Current experimental constraint on $A_-$ has been shown in eqs. (\ref{A-}) and (\ref{A-1}). Using the same numerical inputs for ${\cal C}_{10}$ and $|V_{tb}^*V_{ts}|$ as in Ref. \cite{BBS2018},
together with our estimate of ${\rm Re}~{I}$, we find that, the dispersive long-distance two-photon transition may give rise to the theoretical uncertainty of the branching ratio of $B_s\to\mu^+\mu^-$ decay, which could be up to
\beq 5.3\%\sim8.2\%\;\;\; {\rm for}\;\;\; |A_-|<3.4\times 10^{-4},\eeq or \beq
3.7\%\sim5.8\%\;\;\; {\rm for}\;\;\; |A_-|<2.4\times 10^{-4}.
\eeq

This indicates that quite large uncertainty might be induced from the long-distance contribution, comparable with the uncertainties from $f_{B_s}$ and CKM matrix elements. However, it is very likely that these results are overestimated since, at present $A_-$ is constrained only by the upper limit of ${\cal B}(B_s\to\gamma\gamma)$, and its true value should be smaller once we can fix the branching ratio.
Furthermore, in the present work, we are actually concerned about $A_-$ contributed by the long-distance $B_s\to\gamma\gamma$ transition. Unfortunately, experimental observations cannot separate the long-distance and short-distance contributions, only measure their sum. On the other hand, theoretical predictions of ${\cal B}(B_s\to\gamma\gamma)$ are about $10^{-7}\sim 10^{-6}$, still with large uncertainty, and it was argued in Refs. \cite{LZZ99,BB02} that the long-distance contribution to ${\cal B}(B_s\to\gamma\gamma)$  would be suppressed, which will not exceed a few times $10^{-7}$.  Therefore, now it is unlikely to extract the exact long-distance information on this decay, which is needed in our numerical calculation. Considering the current situation of $B_s\to\gamma\gamma$ decay, here we shall take ${\cal B}(B_s\to\gamma\gamma)_{\rm LD}=1\times 10^{-6}$ and $1\times 10^{-7}$ (LD denoting long-distance), respectively, as examples to illustrate the numerical analysis. Thus the uncertainties in ${\cal B}(B_s\to\mu^+\mu^-)$ could be
\beqn
&&3.0\% \sim 4.6\% \;\;\; {\rm for}\;\;\; |A_-|\gg |A_+|,\\
&&2.1\% \sim 3.3\% \;\;\; {\rm for}\;\;\; |A_-|\sim |A_+|
\eeqn
if ${\cal B}(B_s\to\gamma\gamma)_{\rm LD}=10^{-6}$, and
\beqn
&&0.9\% \sim 1.5\% \;\;\; {\rm for}\;\;\; |A_-|\gg |A_+|,\\
&&0.7\% \sim 1.0\% \;\;\; {\rm for}\;\;\; |A_-|\sim |A_+|
\eeqn
if ${\cal B}(B_s\to\gamma\gamma)_{\rm LD}= 10^{-7}$. These results are still comparable with some theoretical uncertainties discussed in Refs. \cite{BGHMSS, BBS2018}. Hopefully,  future precise measurement and/or theoretical study of the double radiative $B_s$ decay could help to fix the value of $A_-$ or impose more strict constraints on it, which may improve our predictions.

\section{Discussion and summary}

 Rare $B_s\to\mu^+\mu^-$ decay has been observed experimentally. Theoretically, the decay rate is dominated by the short-distance contribution in the SM, which has been calculated very precisely. Thus this muonic decay would provide a very interesting window both to test the SM and to search for new physics. Here we should emphasis that, the main purpose of the present paper is to examine whether the long-distance contribution via $B_s\to\gamma^*\gamma^*\to \mu^+\mu^-$ transition could lead to any significant impact on this decay or not, instead of pursuing an model-independent way to calculate this long-distance contribution, since the latter is a very difficult even impossible task now.  Our study, with some model-dependent assumptions, indicates that it can give rise to new theoretical uncertainty in the branching ratio of $B_s\to\mu^+\mu^-$. This seems not good news because this uncertainty might obscure the new physics signal if the signal is not large.

As mentioned in the section of Introduction, a power-enhanced NLO electromagnetic correction to $B_q\to\ell^+\ell^-$ decay has been found in Ref. \cite{BBS2018}. It is seen that, from the second and third diagrams of Fig. 1 in Ref. \cite{BBS2018}, the two-photon intermediate state also plays some roles. However, those results cannot easily be compared with ours since their diagrams are basically at the quark-level while our calculation has been done mostly at the hadronic level. Comparing Fig. \ref{figure1} in our paper with their two-photon diagrams, one may note that the absorptive part amplitudes, given by the on-shell two-photon exchange, could have some overlaps in these two calculations. This is however no matter since the two-photon contribution alone is very small, we are actually concerned about the dispersive part amplitude and its interference with the short-distance one. In our approach to compute the dispersive two-photon amplitude, the hadronic $B_s\gamma^*\gamma^*$ form factor plays a vital role, and currently we have to adopt some models to formulate it. One cannot expect that these long-distance effects have been included in Ref. \cite{BBS2018}.

To summarize, we have investigated the dispersive contribution of the two-photon intermediate state to the decay $B_s\to\mu^+\mu^-$. The present analysis shows that current experimental data allow the relative large theoretical uncertainty, which arises, in ${\cal B}(B_s\to\mu^+\mu^-)$, from the interference between the long-distance dispersive two-photon amplitude and the dominant short-distance amplitude. The future precise experimental and theoretical studies of the double radiative $B_s$ decay may help to reduce the uncertainty and thus improve our prediction. This novel effect could impact on the branching ratio of $B_s\to\mu^+\mu^-$ decay, which would be essential in interpreting future experimental finds in terms of the SM or new physics scenarios beyond the SM.

\vspace{0.5cm}
\section*{Acknowledgments}
This work was supported in part by the NSF of China under Grant No. 11575175 and by the CAS Center for Excellence in Particle Physics (CCEPP).

\end{document}